\documentclass[a4paper,authoryear,final,times,twocolumn]{elsarticle}
\usepackage{graphicx}
\usepackage{txfonts}
\usepackage{amssymb}

\journal{New Astronomy}

\begin{document}

\begin{frontmatter}

\title{Theoretical cosmic Type Ia supernova rates}

\author[rv]{R. Valiante}
\ead{rosa@oats.inaf.it; valiante@arcetri.astro.it}
\address[rv]{Dipartimento di Astronomia, Universita' di Trieste, via G.B. Tiepolo 11, I-34131}

\author[rv,in]{F. Matteucci}

\author[in]{S. Recchi}

\author[rv]{F. Calura}

\address[in]{I.N.A.F. Osservatorio Astronomico di Trieste, via G.B. Tiepolo 11, I-34131}

\begin{abstract}
The purpose of this work is the computation of the cosmic Type Ia supernova rates, namely the frequency of Type Ia supernovae per unit time in a unitary volume of the Universe. Our main goal in this work is to predict the Type Ia supernova rates at very high redshifts and to check whether it is possible to select the best delay time distribution model, on the basis of the available observations of Type Ia supernovae. We compute the cosmic Type Ia supernova rates in different scenarios for galaxy formation and predict the expected number of explosions at high redshift ($z\geq 2$). Moreover, we adopt various progenitor models in order to compute the Type Ia supernova rate in typical elliptical galaxies of initial luminous masses of $10^{10}$M$_{\odot}$, $10^{11}$M$_{\odot}$ and  $10^{12}$M$_{\odot}$, and compute the total amount of iron produced by Type Ia supernovae in each case. In this analysis we assume that Type Ia supernovae are caused by thermonuclear explosions of C-O white dwarfs in binary systems and we consider the most popular frameworks: the single degenerate and the double degenerate scenarios. The two competing schemes for the galaxy formation, namely the monolithic collapse and the hierarchical clustering, are also taken into account, by considering the histories of star formation increasing and decreasing with redshift, respectively. We calculate the Type Ia supernova rates through an analytical formulation which rests upon the definition of the SN Ia rate following an instantaneous burst of star formation as a function of the time elapsed from the birth of the progenitor system to its explosion as a Type Ia supernova (i.e. the delay time). What emerges from this work is that: \emph{i)} we confirm the result of previous papers that it is not easy to select the best delay time distribution scenario from the observational data and this is because the cosmic star formation rate dominates over the distribution function of the delay times; \emph{ii)} the monolithic collapse scenario for galaxy formation predicts an increasing trend of the SN Ia rate at high redshifts (mainly due to the contribution by massive spheroids), whereas the predicted rate in the framework of a decreasing cosmic star formation rate, more in agreement with the hierarchical scenario, drops dramatically at high redshift; \emph{iii)} for the elliptical galaxies we note that the predicted maximum of the Type Ia supernova rate depends on the initial galactic mass. The maximum occurs earlier (at about 0.3 Gyr) in the most massive ellipticals, as a consequence of the assumed downsizing in star formation. In addition, we find that the Type Ia supernova rate per unit mass at the present time is higher in bluer ellipticals (i.e. the less massive ones).
\end{abstract}


\end{frontmatter}

\twocolumn

\section{Introduction}
The supernovae (SNe) of Type Ia are fundamental for understanding a number of astrophysical problems of primary importance, such as \emph{i)} the SN progenitors, \emph{ii)} the determination of cosmological parameters, \emph{iii)} the chemical enrichment of galaxies and \emph{iv)} the thermal history of the interstellar (ISM) and intracluster medium (ICM). The evolution of the rate of Type Ia SNe with cosmic time is a fundamental ingredient for the study of all these issues. The observed features of SNe Ia suggest that the majority of these objects may originate from the thermonuclear explosion of a C-O white dwarf (WD) of mass $\sim 1.4$ M$_{\odot}$ (Chandrasekhar mass) in binary systems (Chandra exploders). So each SN Ia should be the result of the explosion of the same mass. However, Phillips (1993) pointed out that there is a significant intrinsic dispersion in the absolute magnitudes at maximum light of local Type Ia SNe. This result was interpreted to arise from a possible range of masses of the progenitors or from variations of the explosion mechanism. \\ 
Here we will consider only Chandra exploders for which two scenarios have been proposed:
a) The \emph{Single Degenerate} (SD) scenario, i.e. the accretion of matter via mass transfer from a non-degenerate companion, a red giant or a main sequence star (e.g. Whelan \& Iben 1973). In this scenario the mass range for the secondary components of the binary system is $0.8-8$M$_{\odot}$, while the primary masses should be in the range $2-8M_{\odot}$. The upper limit is given by the fact that stars with masses $M>8M_{\odot}$ ignite carbon in a non-degenerate core and do not end their lives as C-O WDs. The lower limit is instead obviously due to the fact that we are only interested in systems which can produce a Type Ia SN in a Hubble time. The clock to the explosion is given by the lifetime of the secondary component.
b) The \emph{Double Degenerate} (DD) scenario, i.e. the merging of two C-O WDs which reach the Chandrasekhar mass and explode by C-deflagration (e.g. Iben \& Tutukov 1984). The merging is due to the loss of orbital angular momentum due to gravitational wave radiation. In the Iben \& Tutukov paper, the progenitor masses were defined in the range 5-8M$_{\odot}$ to ensure two WDs of $\sim 0.7$M$_{\odot}$ and then reach the Chandrasekhar mass. The clock for the explosion in this model is given by the lifetime of the secondary star plus the time necessary to merge the system due to gravitational wave radiation. 
This scenario requires the formation of two degenerate C-O WDs at an initial separation less than $\sim 3 R_{\odot}$ and this can occur by means of two different precursor systems: a close binary, and a wide binary. 
The two scenarios differ for the efficiency of the common envelope phase during the first mass transfer, and therefore for the separation attained at the end of the first common envelope phase.

Different arguments can be found in favor or against both SD and DD scenarios and the issue of DD vs. SD is still debated (e.g. Branch et al. 1995, Napiwotzki et al. 2002, Belczynski et al. 2005). The Type Ia SN rate is the convolution of the distribution of the explosion times, usually called the time delay distribution function (DTD), with the star formation history. Several attempts of comparing the Type Ia SN rate and the cosmic star formation rate evolution with redshift have already appeared (e.g. Gal-Yam \& Maoz, 2004; Dahlen et al. 2004; Cappellaro et al. 2005; Neill et al. 2006; Forster et al. 2006; Barris \& Tonry, 2006;  Poznansky et al. 2007; Botticella et al. 2008, Blanc \& Greggio, 2008; among others), but  no clear conclusions arised yet on this point.\\ 
In this paper, besides the above mentioned scenarios for Type Ia SN progenitors we will test the empirical DTD suggested by Mannucci et al. (2006),together with different star formation histories. In particular, we will study the Type Ia SN rate in ellipticals and different cosmic star formation rates, including a strongly increasing cosmic star formation with redshift, based on the monolithic scenario of galaxy formation and never considered in previous studies.\\
The paper is organized as follows: in Section 2 the adopted formulation for the SN Ia rate is presented, together with the description of the adopted star formation histories. In section 3 the models for the distribution function of the delay times are presented and in section 4 the predicted SN Ia rates are discussed and compared with the availale data. Finally, in section 5 some conclusions are drawn. 
    
\section{The computation of the Type Ia supernova rates}
From a theoretical point of view, the Type Ia SN rate is difficult to derive, because the nature of the progenitors of SNIa events is still an open question.\\
One of the first model proposed for the calculation of the Type Ia SN rates was introduced by Greggio \& Renzini (1983, hereafter GR83). This formulation is based on the Whelan and Iben (1973) model (SD scenario). 
Greggio (1996) revised the computation of the SN Ia rate in the framework of the SD model and suggested a more detailed criterion for the formation of a system which can eventually explode as Type Ia SN. 
Another model was proposed by Kobayashi et al. (1998) and Kobayashi, Tsujimoto \& Nomoto (2000). They adopted the Hachisu et al. (1996; 1999) model and considered two possible progenitor systems: either a WD plus a red giant (RG) star or a WD plus a main sequence (MS) star, plus a metallicity effect delaying the formation of  systems which could give rise to SNe Ia. 
Matteucci \& Recchi (2001, hereafter MR01) calculated the Type Ia SN rate for different star formation histories in galaxies in the framework of the SD scenario. They concluded that the best prescriptions to obtain SN Ia rates in agreement with the observations seems to be those of GR83 (with respect to the approach developed by Greggio 1996 and Kobayashi et al. 1998). Tornamb\'e \& Matteucci (1986) formulated a SN Ia rate in the DD scenario and applied it to galactic chemical evolution models. In the past years, many other authors (e.g. Ruiz-Lapuente \& Canal, 1998; Madau, Della Valle \& Panagia, 1998; Dahlen \& Fransson, 1999; Sadat et al., 1998; Yungelson \& Livio, 2000, Han \& Podsiadlowski, 2004; Scannapieco \& Bildsten, 2005; Hachisu, Kato \& Nomoto, 2008) presented formulations for the computation of the rate of Type Ia SNe, alternative to that of GR83. Basically, most of them computed the SN Ia rate by introducing the function describing the delay time between the formation of the SN progenitors and its explosion (i.e. DTD). One of the most recent formulation is that of Greggio (2005, hereafter G05) which will be adopted here.

\subsection{The formulation of the SN Ia rate of G05}
 Following the G05 formalism, at a given epoch \emph{t} the SN Ia rate is:
\begin{equation}
   R_{Ia}(t) = k_{\alpha} \int_{\tau_{i}}^{min(t,\tau_{x})} A_{Ia}(t-\tau) \psi(t-\tau) DTD(\tau)d\tau,
\end{equation}
where $\tau$ is the delay time, 
defined in the range ($\tau_{i}$, $\tau_{x}$), which are the minimum and maximum possible delay times ( i.e. the lifetimes of the secondaries in the SD model and the lifetimes of the secondaries plus the gravitational time delays in the DD model), so that:
\begin{equation}
 \int_{\tau_{i}}^{\tau_{x}} DTD(\tau) d\tau =1.
\end{equation} 
The constant $k_{\alpha}$ is the number of stars per unit mass in one stellar generation and depends on the initial mass function (IMF):
\begin{equation}
   k_{\alpha} = \int_{0.1M_{\odot}}^{100M_{\odot}} \phi(M) dM.
\end{equation}
being $\phi(M)$ the chosen IMF, given by:
\begin{equation}
   \phi(M) = C M^{-(1+x)},
\end{equation}
where C is the normalization constant derived from:
\begin{equation}
   \int_{0.1M_{\odot}}^{100\rm{M}_{\odot}} M \phi (M) dM = 1.
\end{equation}
In this work we adopt a Salpeter IMF, for which $k_{\alpha}$ is equal to 2.83. $A_{Ia}(t-\tau)$ is the fraction of binary systems which give rise to Type Ia SNe relative to the whole range of star masses ($0.1-100 M_{\odot}$). In the formulation proposed by GR83 and MR01 the realization probability $A_{Ia}$ is instead relative to the mass range $3-16 M_{\odot}$ (i.e. the minimum and maximum total mass of the binary system). In Eq. (1) $A_{Ia}$ is a function of the epoch at which the stellar generation is born ($t-\tau$). However, in this work, this quantity is treated as a constant and it is chosen to reproduce the present day SN Ia rate observed value. This formulation shows that the basic ingredients for computing the SN Ia rates are the DTD and the star formation rate (SFR), $\psi$.  

For the analysis described in this paper we consider the SFR of typical elliptical galaxies and the cosmic SFR, namely the SFR in a unit comoving volume of the Universe containing galaxies of all morphological types. These functions are described in the following sections.
\subsection{The star formation rate in elliptical galaxies}
The SFR, i.e. the amount of gas turning into stars per unit time, is usually expressed in M$_{\odot}$ yr$^{-1}$. In the case of elliptical galaxies we adopted the model of chemical evolution of Pipino \& Matteucci (2004). The model is based on the assumption that ellipticals form by means of a rapid collapse of pristine gas, thus producing very high rates of SF (starburst-like regime). The SF is halted as the energy of the ISM, heated by stellar winds and SN explosions, exceeds the binding energy of the gas, thus producing a galactic wind. This happens on a timescale always less than 1 Gyr, varying with the galactic mass. After the development of this wind, no star formation is assumed to take place. The SN feedback is taken into account together with the cooling of SN remnants. Massive but diffuse halos of dark matter around the galaxies are considered. The Salpeter IMF, considered constant in space and time, is adopted. The SFR has a simple form, given by the Schmidt (1959) law,
with an efficiency of star formation which is higher in more massive objects which evolve faster than less massive ones (inverse-wind scenario by Matteucci 1994, also called \emph{downsizing}). 
The reason for this choice is that an increasing efficiency of star formation with mass reproduces very well the observed increase of [Mg/Fe] versus Mass in ellipticals. Therefore, the efficiency of star formation has been calibrated to reproduce such a relation (see Pipino \& Matteucci 2004 for details).  As it can be seen in Fig. 1, we consider the specific SFRs (SSFR, i.e. the SFR per unit mass) corresponding to typical ellipticals of $10^{10}$, $10^{11}$, and $10^{12} \rm{M}_{\odot}$ of luminous mass. These histories of SF will then be used in the next chapter to predict the SN Ia rate under different assumptions about the DTD function.
\begin{figure}
  \centering
  \includegraphics[width=8.3cm]{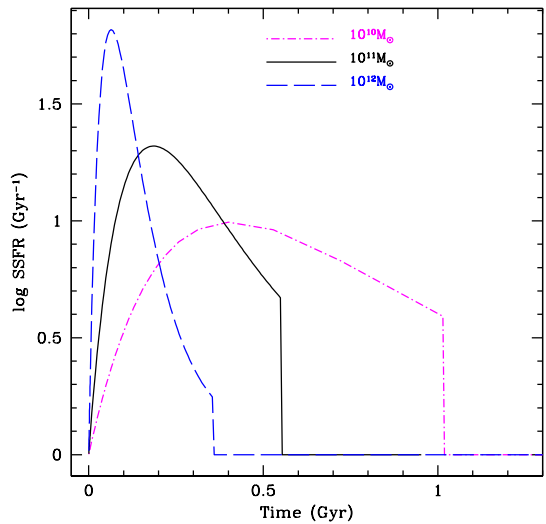}
   \caption{The SFR per unit mass of typical elliptical galaxies of different initial luminous masses, in units of Gyr$^{-1}$, taken from Pipino \& Matteucci (2004). Magenta dot-dashed line: a model for $10^{10} \rm{M}_{\odot}$. Black solid line: a model for $10^{11} \rm{M}_{\odot}$. Blue dashed line: a model for $10^{12} \rm{M}_{\odot}$. Note that the most massive ellipticals stop to form stars before the less massive ones. \, \, \, \, \, \, \, \, \, \, \, \, \, \, \, \, \, \, \, \, \, \, \, \, \, \, \, \, \, \, \,\,\,\,\,\,\,\, \, \, \, \, \, \, \, \, \, \, \, \, \, \, \, \, \, \, \, \, \, \, \, \, \, \, \, \, \,  \label{SFREll}}
\end{figure}
\subsection{The cosmic star formation rates}
The cosmic SFR is usually expressed in M$_{\odot}$yr$^{-1}$Mpc$^{-3}$, in other words it is a SFR density (SFRd) in a unitary volume of the universe. Actually, the cosmic star formation history is measured up to z$\sim$6 (see e.g. Hopkins, 2004; Hopkins \& Beacom, 2006) thanks to the observations at UV, sub-mm and FIR wavelengths, although the error bars at high redshift are huge and the rates still uncertain.
In the framework of a Lambda cold dark matter cosmology ($\Lambda CDM$) with $\Omega_{M}=0.3$, $\Omega_{\Lambda}=0.7$ and $H_{0}\simeq 70$km s$^{-1}$Mpc$^{-3}$ we have selected five  different cases for the SFRd, as proposed in the literature.
The models are those proposed by Calura Matteucci \& Menci (2004, hereafter CMM04), whereas the others are from Madau, Della Valle \& Panagia (1998, hereafter MDP98), Strolger et al. (2004, hereafter S04) and Cole et al. (2001) and they are the best fits to various observations. It is worth noting that some of them predict a decrease of the SFRd at high redshift, in agreement with the hierarchical clustering scenario, but not supported by the present data indicating a rather constant behaviour (see fig. 2). In particular, classical hierarchical clustering (HC) models for galaxy formation predict that ellipticals form continuously from the mergers of bulge-disk systems or other ellipticals, and that most galaxies never experience SFRs in excess of a few solar masses per year. On the other hand, in the monolithic scenario the spheroids form first and the SFRs in massive systems can be as high as thousands solar masses per year. CMM04 computed the theoretical cosmic SFRd as a function of redshift 
by means of detailed chemical evolution models for galaxies of different morphological types (Ellipticals, Spirals and Irregulars). Detailed descriptions of the chemical evolution models can be found in the works of Matteucci (1994) and Pipino \& Matteucci (2004) for ellipticals (as described in sect. 2.2), Chiappini, Matteucci \& Romano (2001) for spirals, and Bradamante, Matteucci, \& D'Ercole (1998) for irregulars. 

The cosmic SFR density was then computed by CMM04 according to:
\begin{equation}
    \dot{\rho}_{\star}(z) = \sum_{i} \rho_{B,i}(z) (\frac{M}{L})_{B,i} \psi_{i}(z),
\end{equation}
where $\rho_{B,i}$ represents the B-band luminosity density (LD), $(\frac{M}{L})_{B,i}$ is the B-band mass-to-light ratio and $\psi_{i}$ is the SFR for the galaxies of the \emph{i}th morphological type. 
Therefore, in the CMM04 model, the total comoving SFR density is given by the sum of the contributions of all the different morphological types ( see Calura \& Matteucci, 2003 for details) and predicts a peak at the redshift of galaxy formation due to starbursts in spheroids (see Fig. 2; dot-dashed line). 

It is worth noting that the CMM04 cosmic rate is marginally consistent with the data for redshifts $z>3$, and this is because it represents a real prediction, whereas the other rates are best fits to the data. On the other hand, the error bars of the data are very large especially at high redshift and it is not possible to distinguish whether the SFRd decreases, stays constant or increases for redshifts larger than 4.  \\

MDP98 presented two cosmic SFRd models (hereafter MDP1 and MDP2), expressed by two analytical fits, $\dot{\rho}_{MDP_{1}}$ and $\dot{\rho}_{MDP_{2}}$:
\begin{equation}
    \dot{\rho}_{MDP_{1}}(t)=a_{1}[t_{9}^{a_{2}}e^{-t_{9}/a_{3}}+a_{4}(1-e^{-t_{9}/a_{3}})] 
\end{equation}
and
\begin{equation}
    \dot{\rho}_{MDP_{2}}(t)=a_{1}e^{-t_{9}/a_{6}}+a_{4}(1-e^{-t_{9}/a_{3}})+a_{5}t_{9}^{a_{2}}e^{-t_{9}/a_{3}}  
\end{equation}
in M$_{\odot}$ yr$^{-1}$ Mpc$^{-3}$. The time $t_{9}$ is the Hubble time at redshift z, given by $t_{9}=13(1+z)^{-3/2)}$, in Gyr and the values for the coefficients are given in Table 1.
The MDP1 model (Eq. 7, short-dashed line in Fig. 2) was built in order to fit the evolution of the observed comoving luminosity density at that time. The MDP2 model (Eq. 8, long-dashed line in Fig. 2) instead gives a larger SFRd at high redshifts and is designed to mimic the MC scenario (as in the model of CMM04). The last analytical model we have considered is that computed by S04, assuming a modified version of the parametric form of the SFR as suggested by Madau et al. (1998), and it takes the correction for extinction into account (see Fig. 2; dotted line). It is expressed in M$_{\odot}$ yr$^{-1}$ Mpc$^{-3}$ and is given by:
\begin{equation}
\dot{\rho}_{S04}(t)=a_{1}(t^{a_{2}}e^{-t/a_{3}}+a_{4}e^{d(t-t_{0})/a_{3}})
\end{equation}
where $t_{0}=13.47 Gyr \,$ is the age of the Universe in Gyr
(corresponding to $z=0$), and the values of the other parameters are
given in Tab. 1. In Fig. 2 all the described models are plotted
together as a function of redshift assuming two different epochs of
galaxy formation $z_{f}= 6, 10$.  An epoch of galaxy formation as
large as $z_f = 10$ is justified by the fact that a number of well
formed massive galaxies has been observed at $z >$ 5 (e.g. Mobasher et
al. 2005; Vanzella et al. 2008).  The conversion $t-z$ is made by
adopting the $\Lambda CDM$ cosmology, as defined before. 
The models are compared with the data provided by
Hopkins (2004) and the parametric form introduced by Cole et
al. (2001) and used by Hopkins \& Beacom (2006) to fit their SFR
measurements (Fig. 2, solid line): $\dot{\rho}_{\star}(z) =
(a+bz)h/(1+(z/c)^{d})$, here with h=0.7. For a Salpeter IMF, the fit
to the data gives $a=0.01334, b=0.175, c=2.93, d=3.01$ (see Blanc \&
Greggio, 2008).

\begin{table*}
  \begin {center}
  \begin{tabular}{|c|c|c|c|c|c|c|}
        \hline
         \textbf{SFRd} & \textbf{$a_{1}$} & \textbf{$a_{2}$} & \textbf{$a_{3}$} & \textbf{$a_{4}$} & \textbf{$a_{5}$} & \textbf{$a_{6}$} \\
         \hline
         $\dot{\rho}_{MDP_{1}}$ & 0.049 & 5 & 0.64 & 0.2 & - & - \\
         \hline
         $\dot{\rho}_{MDP_{2}}$ & 0.336 & 5 & 0.64 & 0.0074 & 0.00197 & 1.6 \\
         \hline
         $\dot{\rho}_{S04}$ & 0.182 & 1.260 & 1.865 & 0.071 & - & - \\
         \hline
   \end{tabular}
   \end{center}
   \caption{Values of the coefficients for the analytical SFR densities. In the first column are indicated the models from Madau Della Valle \& Panagia (1998), $\dot{\rho}_{MDP1}$ and $\dot{\rho}_{MDP2}$, and the one from Strolger et al. (2004), $\dot{\rho}_{S04}$.  \, \, \, \, \, \, \, \, \, \, \, \, \, \, \, \, \, \, \, \, \, \, \, \, \, \, \, \, \, \, \, \, \, \, \, \, \, \, \, \, \, \, \, \, \, \, \, \, \, \, \, \, \, \, \, \, \, \, \, \, \, \, \, \, \, \, \, \, \, \, \, \, \,\, \, \label{tab:Table1}}
\end{table*}
\begin{figure*}
  \centering
  \includegraphics[width=8.3cm]{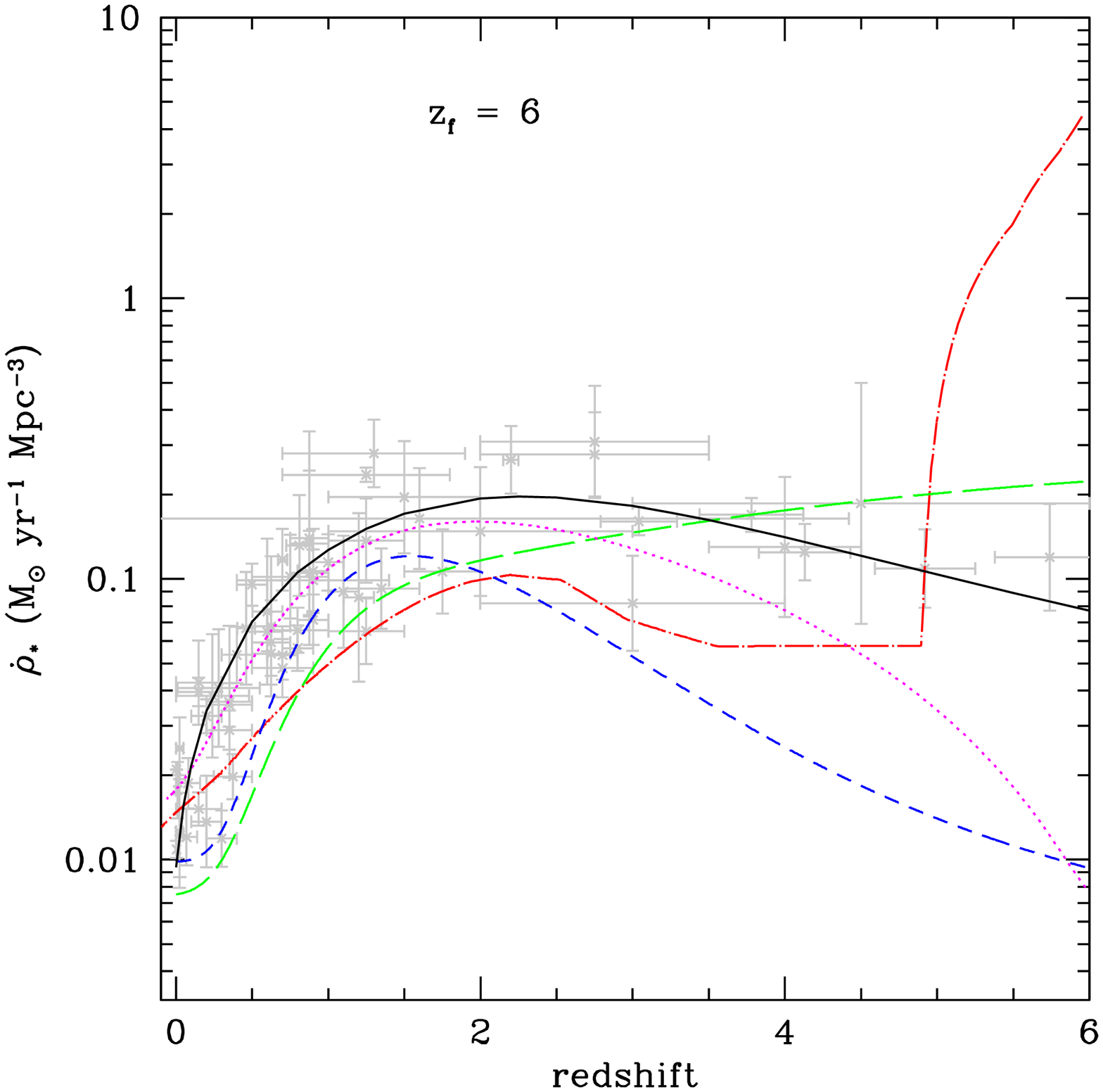}
  \includegraphics[width=8.3cm]{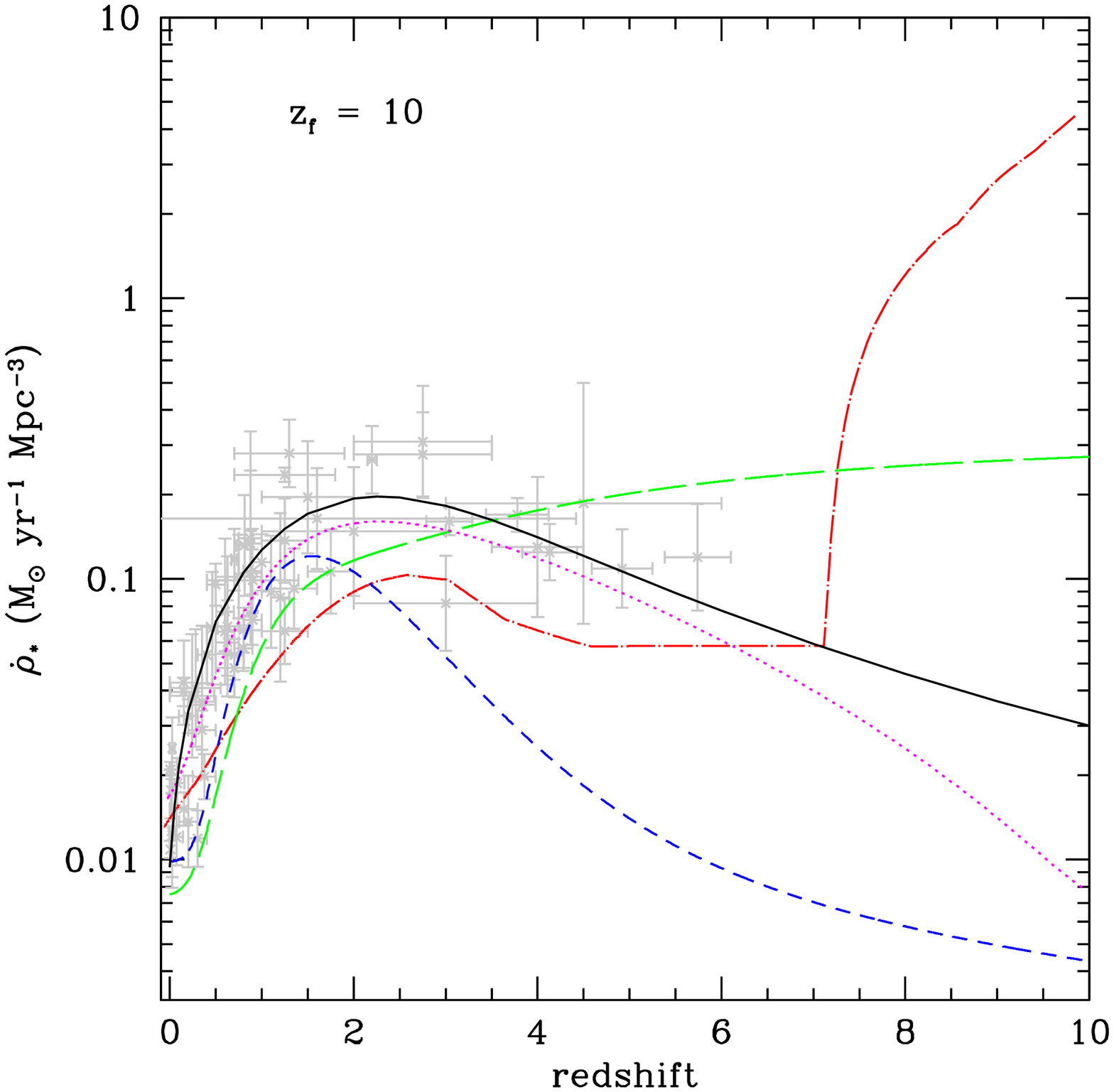}
  \caption{Cosmic SFR densities as functions of redshift, being $z_{f}=6$ (upper panel) and $z_{f}=10$ (lower panel) the galaxy formation epoch. The SFR densities are in units of M$_{\odot}$ yr$^{-1}$ Mpc$^{-3}$. The models are from: Calura, Matteucci \& Menci (2004) (dashed-dotted line); Madau, Della Valle \& Panagia (1998) (short-dashed line); Madau, Della Valle \& Panagia (1998) (long-dashed line); Strolger et al. (2004) (dotted line). The different behaviours of the cosmic SFR are clearly visible toward higher redshifts. The data are from the compilation of Hopkins (2004) and the solid line is the best fit function to the data in the figure by Cole et al. (2001). This latter curve does not depend on the assumed galaxy formation epoch. \, \, \, \, \, \, \, \, \, \, \, \, \, \, \, \, \, \, \, \, \, \, \, \, \, \, \, \, \, \, \, \, \, \, \, \, \, \, \, \, \, \, \, \, \, \, \, \, \, \, \, \, \, \, \, \, \, \, \, \, \, \, \, \, \, \, \, \, \, \, \, \, \, \, \,\label{SFRd}}
\end{figure*}
\section{The distribution functions of the delay times (DTDs)}

For this work we have selected the formulation of MR01 for the SD scenario (see also Matteucci et al. 2006, hereafter M06), that of G05 for the DD wide and close scenarios and the empirical DTD proposed by Mannucci Della Valle \& Panagia (2006, hereafter MVP06).
Different explosion models predict different delay times between the formation of the progenitor system and the SN explosion. In particular, the clock for the explosion in the case of the SD model (GR83 and MR01) is given by the lifetime of the secondary star, ranging between $\sim$0.03-0.04 Gyr (i.e. the lifetime of a $8 M_{\odot}$ star) and 10 Gyr (i.e. the lifetime of a $0.8 M_{\odot}$ star. 
This ensures that this model is able to predict a different from zero present time SN Ia rate for those galaxies where the star formation must have stopped several Gyr ago, such as ellipticals. 
In the DD scenario, the delay time is given by the sum of the lifetime of the secondary plus the gravitational time delay (i.e. the time needed by the two stars to merge after loss of angular momentum due to gravitational wave emission, see Landau \& Lifschitz 1962). In the DD scheme, the minimum delay time is given by the sum of the minimum nuclear delay time (0.03-0.04 Gyr) and the minimum gravitational delay time. This minimum gravitational timescale was suggested to be about 1 Myr by G05.
Mannucci et al. (2005, 2006) suggested, on the basis of observational arguments, that there are two populations of the SN Ia progenitors, where a percentage from 35\% to 50\% of the total SNe Ia explode soon after their stellar birth, namely inside $\sim 0.1$ Gyr (the so-called \emph{prompt} SNe Ia), while the rest has a much wider distribution of lifetimes, exploding over a long period of time (the so-called \emph{tardy} SNe Ia).
M06 tested this hypothesis in models of chemical evolution of galaxies of different morphological type (ellipticals, spirals and irregulars). They showed that this proposed scenario is compatible also with the main chemical properties of galaxies as long as the fraction of prompt SNe Ia is no more than 35\%.
Recently, a direct measurement of the DTD function has been reported by Totani et al. (2008), on the basis of the faint variable objects detected in the Subaru/XMM-Newton Deep Survey (SXDS). They concluded that the DTD function is inversely proportional to the delay time, i.e. the DTD can be well described by a featureless power law (DTD$\propto \tau^{-n}$, with $n\simeq$1) in a range $\tau=$(0.1-10) Gyr. Such a DTD seems to support the DD scenario because the SD scenario, in some binary synthesis codes, predicts prominent peaks in the DTD at characteristic timescales (see e.g. Ruiz-Lapuente \& Canal, 1998; Yungelson \& Livio, 2000; Belczynski et al. 2005; Meng et al. 2008). However, other SD models predict other  DTD shapes similar to the observed DTD (e.g. G05; M06; Kobayashi, Tsujimoto \& Nomoto, 2000). 

\subsection{The DTD of MVP06}
MVP06 had used three important observational results to derive, on empirical grounds, the DTD between the formation of the progenitor star and its explosion as a SN. The fraction of prompt SN Ia is suggested to be about 50\%. This suggestion was based on the the enhancement of the SN Ia rate in radio-loud early-type galaxies, i.e. the fact that early-type radio-loud galaxies show a strong enhancement of the SN Ia rate with respect to the radio-quiet ones (see also  Della Valle et al. 2005).


The result is a bimodal DTD given by the sum of two distinct functions: a prompt Gaussian centered at $5\cdot 10^{7}$yr and a much slower function, either another Gaussian or an exponentially declining function. An analytical formulation for this DTD is given in the work of M06:\\
\begin{equation}
    \log DTD(\tau)=
    \left\{
    \begin{array}{rl}
      1.4-50(\log\tau-7.7)^2 & \mbox{ for } \tau< 10^{7.93}\\
      -0.8-0.9(\log\tau-8.7)^2 & \mbox{ for } \tau> 10^{7.93}.
     \end{array}
     \right.
\end{equation}
where the time is expressed in years. This new formulation, being analytical, is easy to implement in a galactic chemical evolution code, in particular adopting the formalism developed by G05 (Eq. 1).
As pointed out by M06, in the SD scenario, such a DTD can be justified if one assumes that the function describing the distribution of the mass ratios inside the binary systems, $f(\mu)= 2^{1+\gamma} (1+\gamma) \mu^{\gamma}$ (Tutukov \& Yungelson, 1980), is a multi-slope function.
In this expression, $\mu =M_{2}/M_{B}$ is the mass fraction of the secondary and $f(\mu)$ is normalized to 1 between $\mu=0$ and $\mu=1/2$. In the case of the DTD of MVP06, a slope $\gamma=2.0 \,$ should be assumed for systems with mass in the range $5-8 M_{\odot}$ (i.e. in this mass range system where the primary and secondary mass are almost equal are preferred), whereas a negative slope, $\gamma\simeq -0.8, -0.9$, should be adopted for masses lower than $5 M_{\odot}$ (that means that systems where $M_{1}\gg M_{2}$, in this mass range, are favored).

Following M06, we assume, for this empirical DTD, the prescription of the SD model, thus $\tau_{i}$ and $\tau_{x}$ are the lifetimes of a 8M$_{\odot}$ and 0.8M$_{\odot}$, respectively.
\subsection{The DTD of MR01}
The DTD of MR01 is computed adopting the following formalism:
\begin{equation}
DTD(\tau) \propto \tilde{\phi}(M_{2}) \dot{M_{2}},
\end{equation}
that corresponds to the the SN Ia rate for an instantaneous starburst. The function $\tilde{\phi}(M_{2})$ is the mass function of the secondary component, and in this case is given by:
\begin{equation}
\tilde{\phi}(M_{2})=2^{1+\gamma} (1+\gamma) M_{2}^\gamma (M_{b}^{(-s-\gamma)}-M_{B}^{(-s-\gamma)})/(-s-\gamma),
\end{equation}
with $s=1+x$ ($x$ is the Salpeter index).
The derivative $\dot{M}_{2}=dM_{2}/dt$ was obtained adopting the inverse of the formula (MR01):
\begin{equation}
\tau(M)=10^{(1.338-\sqrt{1.79-0.2232(7.764-\log M)})/0.1116}
\end{equation}
which defines the relation between the stellar mass (M) and the main sequence lifetime ($\tau$), with M expressed in M$_{\odot}$ and $\tau$ in yr. 

Here the parameters are: $\gamma=2$ \,  for a one-slope function $f(\mu)$, \, $s=2.35$, $M_{b}=max(2M_{2},M_{min})$, \, $M_{B}=M_{2}+0.5(M_{max})$ with $M_{min}=3$M$_{\odot}$ and $M_{max}=16$M$_{\odot}$ i.e. the minimum and maximum masses of the system (see MR01, GR83). 
\subsection{The DTD of G05 for the DD scenario}
In order to compute the DTD of the DD model, for both wide and close channels, G05 refers to systems with $2\rm{M}_{\odot}\leq M_{1},M_{2}\leq 8\rm{M}_{\odot}$, from which most double C-O WDs form. Typically, the WD mass of both components ranges between 0.6 and 1.2 $\rm{M}_{\odot}$, so that the lifetime, $\tau_{n}$, ranges between about 1 and 0.03 Gyr. 
The DTD depends on the distributions of both $\tau_{n}$ and $\tau_{gw}$, with early explosions provided by systems with short $\tau_{n}$ and $\tau_{gw}$. The delay time is given by $\tau=\tau_{n}+\tau_{gw}$, with $\tau_{gw}$ ranging between 1Myr and a maximum value that is larger than $\tau-\tau_{n}$ for all $\tau_{n}$ (at least for a total delay time up to the Hubble time) in the wide DD scheme, whereas in the close one it is correlated with $\tau_{n}$ 
The DTDs proposed by G05 for the wide and close DD schemes essentially depend on  the following parameters (we refer the reader to G05 for details): 
\begin{itemize}
 \item the distribution function of the final separations, which follows a power law: $n(A)\propto A^{\beta_{a}}$;
\item the distribution of the gravitational delay times. It  follows a power law, $n(\tau_{gw})\propto \tau_{gw}^{\beta_{g}}$; 


\item  the maximum nuclear delay time, $\tau_{n,x}$. 
G05 treats $\tau_{n,x}$ as a parameter, comparing the results obtained adopting 0.4, 0.6, 1 Gyr corresponding to the lifetimes of a 3, 2.5 and 2 $M_{\odot}$, respectively.
\end{itemize}
Following the results obtained by G05, we assume $\tau_{n,x}=0.4$Gyr, $\beta_{a}=-0.9$ and $\beta_{g}=-0.75$.
Then the DTD is given by: 
\begin{equation}
DTD(\tau) \propto \int_{\tau_{n,i}}^{min(\tau_{n,x},\tau)} n(\tau_{n}) S^{W}(\tau,\tau_{n}) d\tau_{n} \, \, \, \, \, \, \, \, \, \, \, \, \, \, wide DD
\end{equation}
\begin{equation}
DTD(\tau) \propto \int_{\tau_{n,inf}}^{min(\tau_{n,x},\tau)} n(\tau_{n}) S^{C}(\tau,\tau_{n}) d\tau_{n} \, \, \, \, \, \, \, \, \, \, \, \, \, \, close DD
\end{equation}
complemented with $DTD(\tau)=0$ for $\tau\leq \tau_{i}$ and $\tau\geq \tau_{x}$.\\
where $S^{W}$ and $S^{C}$ are functions of $\beta_{a}$ and $\beta_{g}$, respectively (see G05 for details). 
We remark that, in the different models, the number of \emph{prompt} Type Ia SNe   has an important impact in defining the shape of the DTD. In the DTD models proposed by MVP06, MR01 (SD scenario) and G05 (DD scenario) about the 50\%, 13\% and 7\% of the SNe Ia, respectively, are found to be \emph{prompt}.
\section{Results}
By convolving the different DTDs with the various SFRs, described in the previous sections, we obtain the following results. We first concentrate on a particular morphological type, ellipticals and then expand the analysis to the cosmic rates.
\subsection{The predicted Type Ia rates in ellipticals}
We consider elliptical galaxies of different baryonic mass ($10^{10}$, $10^{11}$, $10^{12}$ M$_{\odot}$) adopting the SFRs of Pipino \& Matteucci (2004). The parameter $A_{Ia}$, i.e. the fraction of systems which are able to originate a SN Ia explosion, has been chosen a posteriori to ensure that the predicted present day SNe Ia rate in ellipticals is compatible with the observed average value, given by Cappellaro et al. (1999):  $\, 0.18\pm 0.06$ SNu, being 1SNu$=1SN/10^{10}$L$_{\odot B}/century$. In particular, we have chosen the realization fraction $A_{Ia}$ for each DTD, by reproducing the present time Type Ia SN rate in a typical elliptical of present time luminous mass of about $3.5\cdot 10^{10}$M$_{\odot}$, corresponding to an initial luminous mass of $10^{11}$M$_{\odot}$ . 
To obtain the present time Type Ia SN rate, in units of $SNe ~century^{-1}$, for an elliptical with a stellar mass of  $3.5\cdot 10^{10}$M$_{\odot}$, we multiplyied the Cappellaro et al. (1999) rate by the blue luminosity  
predicted by our photometric model for such a galaxy ($L_B=4 \cdot 10^{9}L_{B \odot}$), thus obtaining a Type Ia SN rate of 0.072 $SNe ~century^{-1}$ (see Table 2).

Then the $A_{Ia}$ obtained for this typical elliptical and a given DTD has been applied to ellipticals of different mass. In particular, Figs. 3, 4 and 5 show the results for an elliptical of initial mass $10^{10} M_{\odot} \,$, $10^{11} M_{\odot} \,$, $10^{12} M_{\odot} \,$, respectively, versus redshift. We assumed $z_{f}=6$ for the galaxy formation epoch. 
It is interesting to note that the maximum in the Type Ia SN rate depends on the galactic mass through the star formation history. In particular, for a galaxy with initial mass of $10^{12}$M$_{\odot}$, which suffers a very strong burst of SF with high efficiency and for a shorter time than less massive galaxies (downsizing in star formation), the maximum of the Type Ia SNe occurs at about 0.3 Gyr, whereas in an elliptical with initial mass of $10^{10}$M$_{\odot}$ the peak is at about 1 Gyr. This particular behaviour of the Type Ia SNR with galactic mass is fundamental to reproduce the increase of the average [Mg/Fe] in ellipticals, with galactic mass (see Matteucci 1994). It is worth noting that it is very difficult to reproduce such a trend in the classical HC framework (see e.g. Thomas et al. 2002). 
\begin{figure}
     \centering
     \includegraphics[width=8cm]{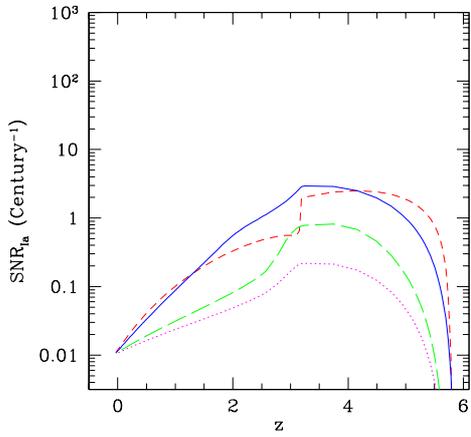}
     \caption{The predicted SN Ia rates for an elliptical galaxy of initial luminous mass of $10^{10}$M$_{\odot}$ versus redshift . The results for different DTD models are shown: MR01 (solid lines) and MVP06 (short-dashed lines) for the SD scenario; G05 for wide (dashed lines) and close (dotted lines) DD scenario. The Type Ia SNR is in unit of century$^{-1}$ and is shown on a logarithmic scale. All the predictions are normalized in order to reproduce the observed present day Type Ia SN rate in ellipticals (see text). Here, the redshift of galaxy formation is assumed to be $z_{f}=6$. \, \, \, \, \, \, \, \, \, \, \, \, \, \, \, \, \, \, \, \, \, \, \, \, \, \, \, \, \, \, \, \, \, \, \, \, \, \, \, \, \, \, \, \, \, \, \, \, \, \label{SNREll10}}
\end{figure}

\begin{figure}
     \centering
     \includegraphics[width=8cm]{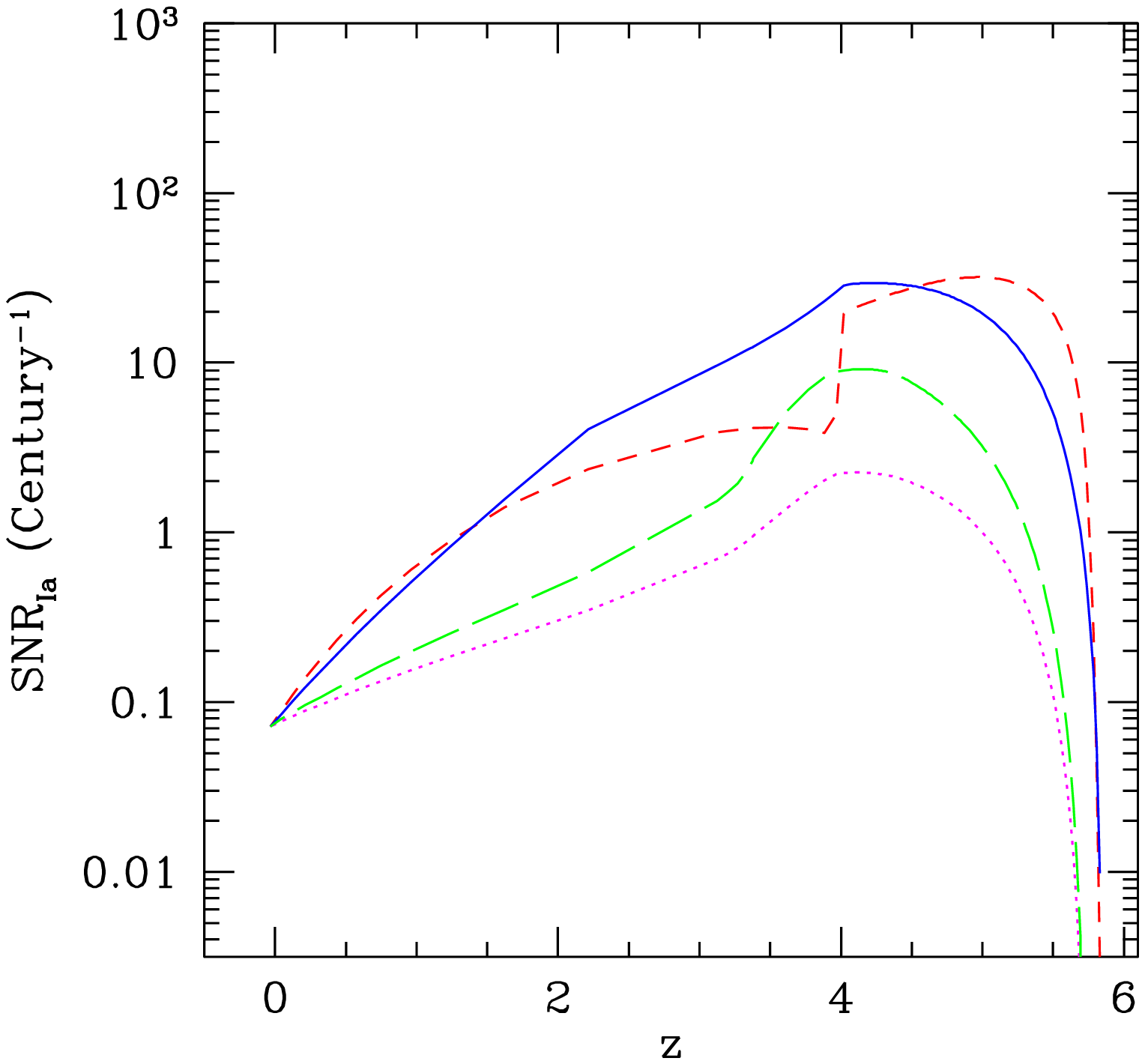}
     \caption{The predicted SN Ia rates for an elliptical galaxy of initial luminous mass of $10^{11}$M$_{\odot}$ versus redshift.  The line types and units are the same as in Fig. 3. \, \, \, \, \, \, \, \, \, \, \, \, \, \, \, \, \, \, \, \, \, \, \, \, \, \, \, \, \, \, \, \, \, \, \, \, \, \, \, \, \, \, \, \, \, \, \, \, \, \, \, \, \, \, \, \, \, \, \, \, \, \, \, \, \, \, \, \, \, \, \, \, \,  \label{SNREll11}}
\end{figure}
\begin{figure}
     \centering
     \includegraphics[width=8cm]{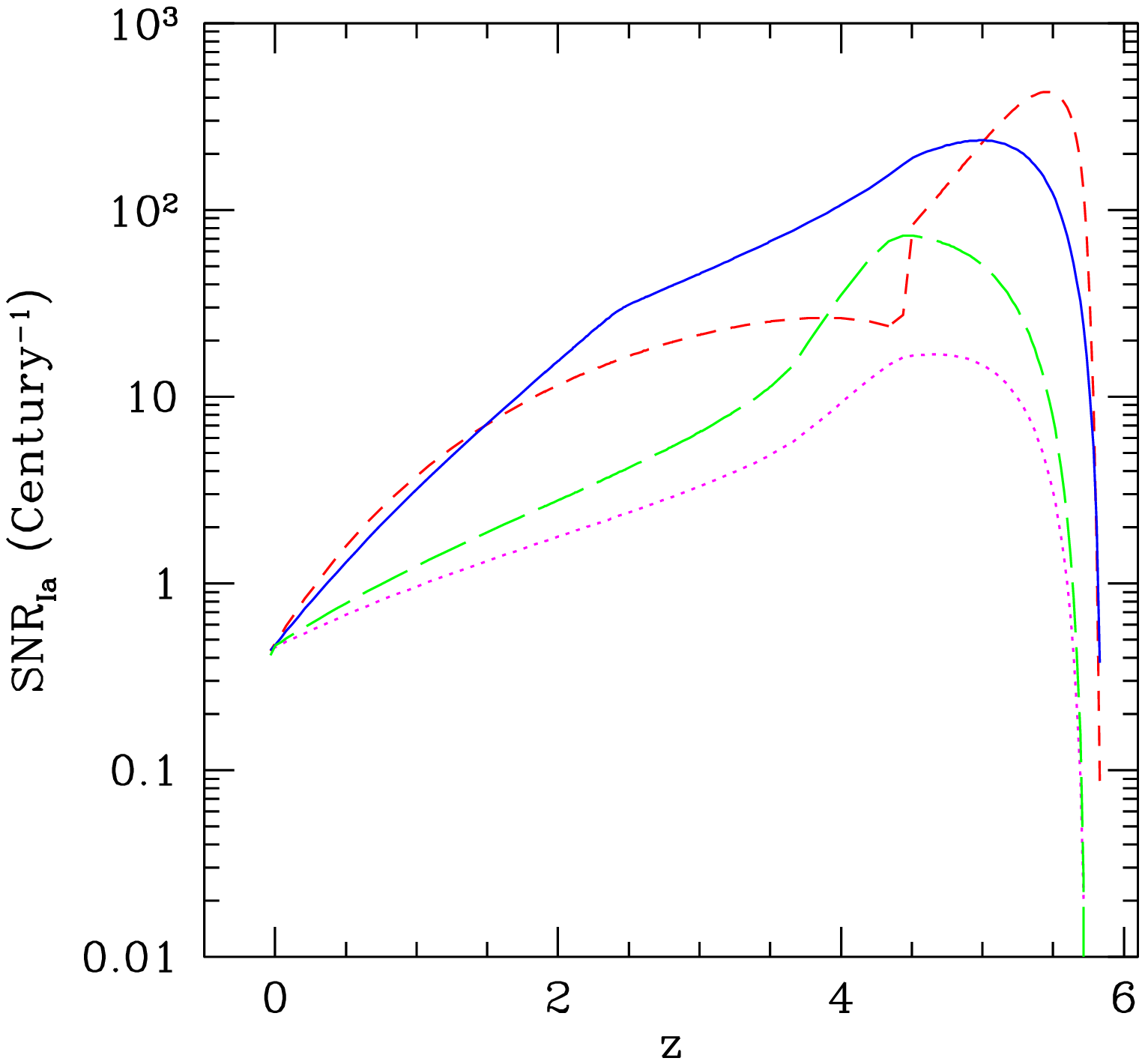}
     \caption{The predicted SN Ia rate for an ellitical galaxy of initial luminous mass of $10^{12}$M$_{\odot}$, versus redshift.  The line types and units are the same as in Fig. 3. \, \, \, \, \, \, \, \, \, \, \, \, \, \, \, \, \, \, \, \, \, \, \, \, \, \, \, \, \, \, \, \, \, \, \, \, \, \, \, \, \, \, \, \, \, \, \, \, \, \, \, \, \, \, \, \, \, \, \, \, \, \, \, \, \, \, \, \, \, \, \, \label{SNREll12}}
\end{figure}

In Table 2 are indicated the total number (SNIa) of SNe Ia explosions over the galactic lifetime, the predicted SN rates (SNR) at the present time, in units of $SNe ~century^{-1}$, and the ejected masses of Fe (after the galactic wind) for typical elliptical galaxies of initial luminous masses $M_{L}$ of $10^{10}$, $10^{11}$, $10^{12} \rm{M}_{\odot} \,$, respectively. The total Fe mass produced by each galaxy was computed assuming that a single SN Ia typically produces about 0.6M$_{\odot}$ of Fe (e.g. Nomoto et al. 1997). For the three different ellipticals, $A_{Ia}=0.0019 \,$, for the MR01 model, $A_{Ia}=0.012 \,$, for MVP06, $A_{Ia}=0.00021 \,$ for close DD and $A_{Ia}=0.0005 \,$ for wide DD are assumed.
The computation of the Fe masses is useful in order to study the enrichment of the intracluster medium.  The predicted values of the Fe masses are in agreement with those predicted by Matteucci \& Vettolani (1988) and by Matteucci \& Gibson (1995). We can note, in all the three cases, that the G05 DD models predict less SNe Ia and less Fe with respect to the SD models. The last analysis we have made on ellipticals is based on the study of the SN Ia rate per unit mass, i.e. expressed in SNuM (Mannucci et al. 2005), being $1 SNuM=1 SN/century/10^{10}$M$_{\odot}$. As suggested by Mannucci et al. (2005), the SN rate normalized to the stellar mass of the parent galaxies contains unique information on the IMF of stars in the range of masses between about 3 and 100 $M_{\odot}$. 
Therefore, it is a very powerful tool for understanding the formation and the chemical evolution of galaxies and constraining their star formation histories. Here we focus the attention on the dependence of the SN Ia rate of elliptical galaxies on their $(B-K)$ color, which is an indicator of the mean age of the population. The SFR is related both to the morphology and to the color of the galaxies (e.g. Kennicutt 1998), but the SFR-color relation is probably tighter than the SFR-morphology one as the color is more directly related to stellar populations than to morphology. Assuming that the stellar mass at the present time of the elliptical galaxies is about $\sim 40\%$ of the initial luminous mass (see Pipino \& Matteucci 2004), the predicted SNRs at the present time, in SNuM (see Figure 6), for the three different galaxies are corresponding to the DTD models from MR01, MVP06 and G05 wide and close DD. The colors have been computed by means of the spectrophotometric code of Jimenez et al. (2004). We can see, in Figure 6, that the bluer is the $(B-K)$ colour of the parent galaxy, the higher is the predicted rate per unit mass at the present time (the first point on the left corresponding to an elliptical of initial luminous mass of $10^{10} M_{\odot}$). This behaviour can be explained by the fact that the peak in the SFR (and hence the maximum in the SN Ia rate) occurs later in the less massive (bluer) ellipticals. 

Unfortunately, the available data give only an average rate of Type Ia SNe as a function of morphological type and not as a function  of colors for ellipticals of different masses (for example SNuM $=(0.027 ^{+0.017}_{- 0.01})$ for $(B-K) > 4.1$ in Mannucci et al. 2005, see also the more recent Mannucci et al. 2008).

\begin{figure}
     \centering
     \includegraphics[width=7.8cm]{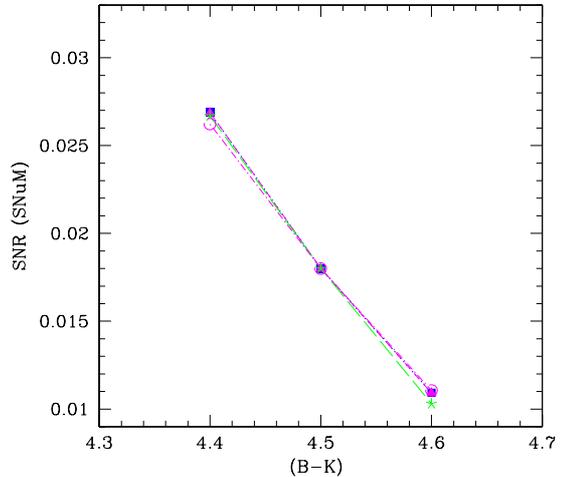}
     \caption{The predicted Type Ia SN rate at the present time vs. $(B-K)$ color of elliptical galaxies. The symbols are the predictions obtained by adopting the DTD from G05 in the wide scheme (stars), the DTD from G05 in the close scheme (open circles), the DTD from MR01 (squares) and the DTD from MVP06 (triangles). The present day SN Ia rate is expressed in SNuM, i.e. in SNe per unit mass per century and it has been computed assuming that the present time stellar mass of the galaxy is about the 40\% of its initial stellar mass. For a typical elliptical of initial luminous mass of $10^{11}$ M$_{\odot}$ we find a SNIa rate of about 0.018 SNuM. The $(B-K)$ color increases with the mass of the galaxy, being reddest for a $10^{12}$M$_{\odot}$.The very small difference between the present time predicted values of the models for the galaxies with initial masses $10^{10}$ and $10^{12}M_{\odot}$ is due to the choice of the realization probability $A_{Ia}$, calibrated to reproduce the present time Type Ia SN rate of a $10^{11}M_{\odot}$.}
 \label{SNRcol}
\end{figure}

\begin{table*}
\begin{center}
\begin{tabular}{|c|c|c|c|}
\hline
 Model & SNIa  &  SNR(century$^{-1}$) & $M_{Fe}$(M$_{\odot}$) \\
\hline 
\hline
\multicolumn{4}{|c|}{M$_{L}=10^{10}$M$_{\odot}$}\\
\hline
 MR01  & $4.33\cdot 10^{7}$  & 0.0108  & $2.6\cdot 10^7$ \\
\hline
 MVP06 & $3.32\cdot 10^{7}$  & 0.0109  & $1.99\cdot 10^7$   \\
\hline
 G05 wide & $1.1\cdot 10^{7}$  & 0.0107  & $0.66\cdot 10^7$   \\
\hline
 G05 close & $0.46\cdot 10^{7}$  & 0.0105  & $0.28\cdot 10^7$   \\
\hline
\multicolumn{4}{|c|}{M$_{L}=10^{11}$M$_{\odot}$}\\
\hline 
 MR01  & $2.7\cdot 10^{8}$  & 0.072  & $1.6\cdot 10^8$ \\
\hline
 MVP06 & $2.23\cdot 10^{8}$  & 0.072  & $1.34\cdot 10^8$   \\
\hline
 G05 wide   & $0.7\cdot 10^{8}$  & 0.072  & $0.42\cdot 10^8$   \\
\hline
 G05 close   & $0.3\cdot 10^{8}$  & 0.072  & $0.18\cdot 10^8$   \\
\hline
\multicolumn{4}{|c|}{M$_{L}=10^{12}$M$_{\odot}$}\\
\hline 
 MR01  & $1.86\cdot 10^{9}$  & 0.437  & $1.12\cdot 10^9$ \\
\hline
 MVP06 & $1.47\cdot 10^{9}$  & 0.434  & $0.88\cdot 10^9$   \\
\hline
 G05 wide   & $0.47\cdot 10^{9}$  & 0.412  & $0.28\cdot 10^9$   \\
\hline
 G05 close   & $0.2\cdot 10^{9}$  & 0.442  & $0.12\cdot 10^9$   \\
\hline
\end{tabular}
\end{center}
\caption{The predictions for elliptical galaxies of initial luminous mass of $10^{10},10^{11}, 10^{12}$ M$_{\odot}$. In the first column the various DTD models are specified, while the second column shows the the total number of SNe Ia exploded over an Hubble time, predicted by each model. The predicted current SN Ia rates (SNR), in units of century$^{-1}$, and the total Fe masses produced, in units of solar masses, are indicated in the third and fourth columns, respectively (see text). \, \, \, \, \, \, \, \, \, \, \, \, \, \, \, \, \, \, \, \, \, \, \, \, \, \, \, \, \, \, \, \, \, \, \, \, \, \, \, \, \, \, \, \, \, \, \, \, \, \, \, \, \, \, \, \, \, \, \, \, \, \, \, \, \, \, \, \, \, \, \, \, \, \, \, \, \, \, \, \, \, \, \,}
\label{tab:tabel}
\end{table*}

\subsection{The predicted cosmic SN Ia rate densities}
We present here the cosmic SN Ia rate densities, namely the number of Type Ia SN explosions per yr and per Mpc$^{3}$, at low and high redshifts. The rates are computed by adopting the SFR densities described before (Fig. 2).
Here we compare the predicted cosmic Type Ia SN rates in order to define the differences due to the choice of different scenarios for the SN Ia progenitors (namely, the SD and DD channels) and of different galaxy formation schemes (namely, the increasing, decreasing and constant SFRd). In all cases the realization probability $A_{Ia}$ was chosen a posteriori in order to reproduce the observed rate of Mannucci et al. (2005, see Table 3) at redshift z$\simeq$0.03. Figs. 7, 8, 9 and 10 illustrate the predicted cosmic Type Ia rates, separately for each DTD, compared with the observed ones at different redshifts, assuming $z_{f}=6$ for the galaxy formation epoch. The values of the parameter $A_{Ia}$ are labeled in each panel, with line types refering to the different SFRds. Some observed Type Ia rates at different redshifts are collected in Table 3.

\begin{table*}[htbp]
\begin{small}
 \begin {center}
  \begin{tabular}{|c|c|c||c|c|c|}
   \hline
    \textbf{z} & \textbf{SN rate} & \textbf{Reference} & \textbf{z} & \textbf{SN rate} & \textbf{Reference}\\ 
           &   \textbf{(10$^{-4}$Mpc$^{-3}$yr$^{-1}$)} &            &   &  \textbf{(10$^{-4}$Mpc$^{-3}$yr$^{-1}$)} & \\
     \hline
      0.03 & 0.28$\pm$ 0.11  & Mannucci et al. 2005 &       0.55 & 0.52$\pm$ 0.14 & Pain et al. 2002\\
     \hline 
      0.10 & 0.32$\pm$ 0.15 & Magdwick et al. 2003 &       0.55$\pm$ 0.05 & 2.04$\pm$ 0.38 & Barris \& Tonry 2006\\
     \hline
      0.11 & 0.37$\pm$ 0.10 & Strolger et al. 2003 &       0.65$\pm$ 0.05 & 1.49$\pm$ 0.31 & Barris \& Tonry 2006\\
     \hline
      0.13 & 0.20$\pm$ 0.08 & Blain et al. 2004 &       0.75$\pm$ 0.05 & 1.78$\pm$ 0.34 & Barris \& Tonry 2006\\
     \hline
      0.2 & 0.189$\pm$ 0.042 & Horesh et al. 2008 &      0.75$\pm$ 0.25 & 0.43$\pm{+0.36 -0.32}$ & Poznanski et al. 2007\\
     \hline
      0.24$\pm$ 0.05 & 0.17$\pm$ 0.17 & Barris \& Tonry 2006 &       0.8 &  1.57$\pm{+0.5 -0.4}$ & Dahlen et al. 2004\\
     \hline 
      0.3 &  0.34$\pm {+0.16 -0.15}$ & Botticella et al. 2008 &       1.2 &  1.15$\pm{+0.7 -0.5}$ & Dahlen et al. 2004\\
     \hline
      0.35$\pm$ 0.05 & 0.53$\pm$  0.24 & Barris \& Tonry 2006 &       1.2$\pm$ 0.2  & 0.75$\pm{+0.35 -0.30}$ & Kuznetsova et al. 2008\\
     \hline
      0.4 &  0.69$\pm {+0.2 -0.4}$ & Dahlen et al. 2004 &       1.25$\pm$ 0.25 & 1.05$\pm{+0.45 -0.56}$ & Poznanski et al. 2007\\
     \hline
      0.4$\pm$ 0.2  & 0.53$\pm{+0.39 -0.17}$ & Kuznetsova et al. 2008 &  1.55$\pm$ 0.15  & 1.2$\pm{+0.58 -0.12}$ & Kuznetsova et al. 2008\\
     \hline
      0.45$\pm$ 0.05 & 0.73$\pm$ 0.24 & Barris \& Tonry 2006 &       1.6 &  0.44$\pm{+0.5 -0.3}$ & Dahlen et al. 2004\\
     \hline
      0.46 & 0.48$\pm$ 0.17 & Tonry et al. 2003 &       1.75$\pm$ 0.25 & 0.81$\pm{+0.79 -0.60}$ & Poznanski et al. 2007\\
     \hline
      0.47 & 0.42$\pm$ 0.06 & Neil et al. 2006 &  &  &  \\
\hline
\end{tabular}
\end{center}
\end{small}
\caption{A compilation of the available observational cosmic Type Ia rates at different redshifts up to z$\sim$1.75. The SN Ia rates are expressed in units of 10$^{-4}$Mpc$^{-3}$yr$^{-1}$ . These data are compared with the predicted cosmic rates in Figs. 7-15}
\label{tab:tabel}
\end{table*}

\begin{figure}
     \centering
     \includegraphics[width=8.5cm]{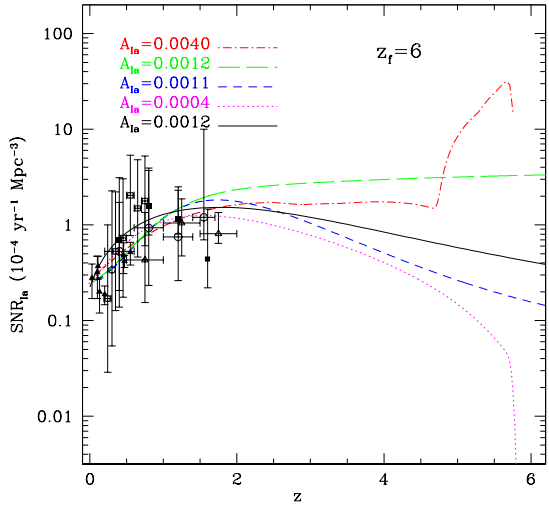}
     \caption{The cosmic SN Ia rate density for the MVP06 DTD. This DTD is convolved with the SFRd from CMM04 (dot-dashed line), MDP2 (long-dashed line), MDP1 (short-dashed line), S04 (dotted line) and Cole et al. (2001, solid line). The short-dashed, dotted and solid curves show a decrease at redshift $z>1$ with respect to the other models (dot-dashed and long-dashed lines). The SN Ia rate is expressed in units of $10^{-4}$yr$^{-1}$Mpc$^{-3}$. The redshift ranges from 0 to 6 (i.e. the epoch at which the star formation is supposed to have started). The solid squares are the data by Dahlen et al. (2004), the open pentagon is by Botticella et al. (2008), the open circles are from Kuznetsova et al. (2008), the open triangles are from Poznanski et al. (2007), the open squares are from Barris \& Tonry (2006), all the other data, given in Table 3, are represented as solid triangles. The predicted present day rates are normalized to reproduce the observed value at z=0 (see text). \, \, \, \, \, \, \, \, \, \, \, \, \, \, \, \, \, \, \, \, \, \, \, \, \, \, \, \, \, \, \, \, \, \, \, \, \, \, \, \, \, \, \, \, \, \, \, \, \, \, \, \, \, \, \, \, \, \, \, \, \, \, \, \, \, \, \, \, \, \, \, \, \, \, \, \label{SNMVP06}}
\end{figure}
\begin{figure}
     \centering
     \includegraphics[width=8.5cm]{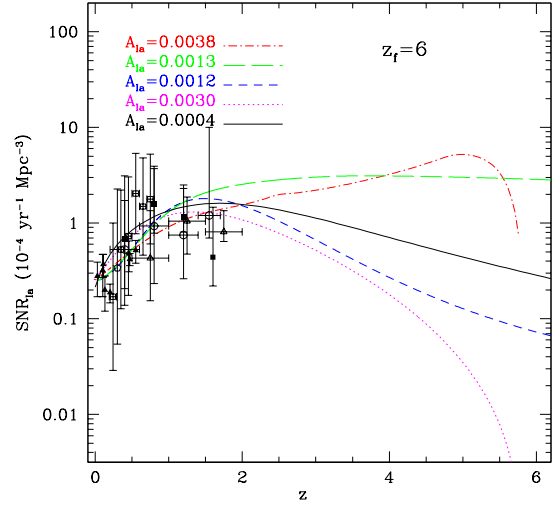}
     \caption{As in Fig. 7, but for the MR01 DTD. \label{SNMR01}}
\end{figure}
\begin{figure}
     \centering
     \includegraphics[width=8.5cm]{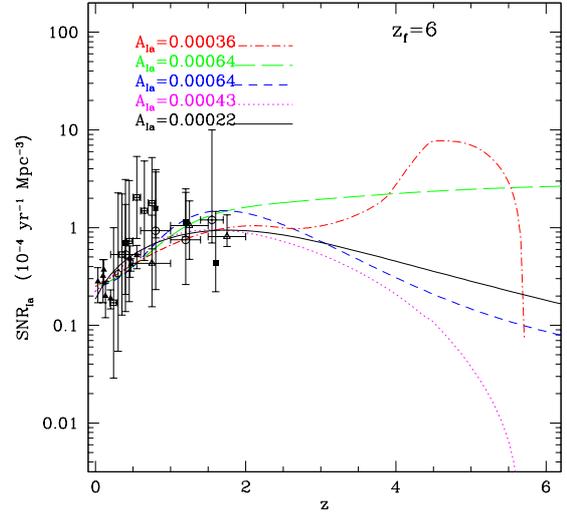}
     \caption{As in Fig. 7, but for the wide DD DTD of G05. \label{SNGR1}}
\end{figure}
\begin{figure}
     \centering
     \includegraphics[width=8.5cm]{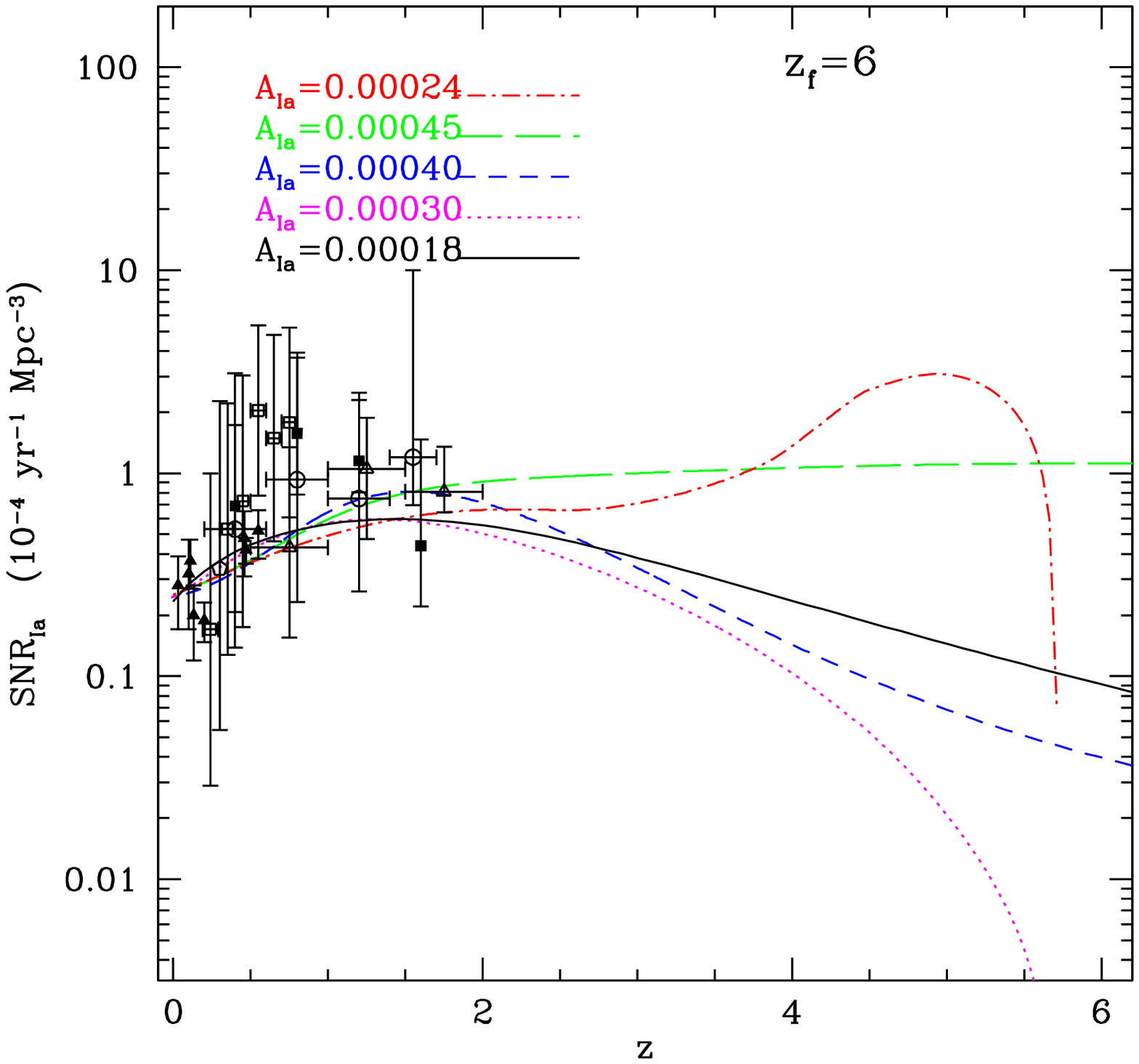}
     \caption{As in Fig. 7, but for the close DD DTD of G05. \label{SNGR2}}
\end{figure}

As it can be seen in these Figures, the predicted rates at low redshift
($z\leq 1 \,$), are similar for all the different DTDs. At higher
redshifts, the rates computed with the SFRds of MDP1 (short-dashed
lines) and Cole et al. (2001, solid lines) 
show a steeper decline than the others, whereas those computed
adopting MDP2 (i.e. for a MC scenario; (long-dashed lines) display a mild
rise up to $z\sim 4 \,$. The most marked differences are shown when
the theoretical SFRd of CMM04 is used. 
In fact, in the scenario of CMM04 the elliptical galaxies form at
very high redshifts and suffer a quite strong star formation, therefore producing
large SN Ia rates at high redshifts. The spirals instead
contribute to the SFRd later and reach a peak at $z\sim 2$. This is a big
difference with the models which follow a hierarchical scheme, in other words with models predicting a decrease of the cosmic SFRd at high redshifts and assume that ellipticals form mainly at lower redshift ($z\sim 1-2$). 
As a consequence of this, the detection of Type Ia SN rates at
high redshift ($z>3$) would help to decide which scenario of galaxy
formation is favoured. In particular, the models adopting the SFR
computed from CMM04 for a MC scenario are the only ones
predicting a very high SN Ia rate at high redshift. The
curve obtained by convolving the bimodal DTD of MVP06  and the SFR of
CMM04, assuming $z_f=6$, shows a high peak at early times (high redshifts), predicting a
larger number of Type Ia SNe ($\sim 73.3\cdot 10^{-4}$ SNe Ia
yr$^{-1}$Mpc$^{-3}$, at $z\sim 4.8$) with respect to the other
models. 
It is important to point out that, at high redshifts ($z>1$), the majority of the models seem
to overestimate the data by Dahlen et al. (2004) solid squares). It has been suggested that these data are very difficult
to reproduce by means of standard SN rate models (Dahlen et al. 2004,
Mannucci et al. 2005), unless a large delay time ($\sim 4$ Gyr)
between the epoch of star formation and the explosion of Type Ia SN is
assumed. This means that one should wait a time comparable to $\sim 4$
Gyr after the beginning of star formation in order to observe a
significant contribution from Type Ia SNe to the chemical enrichment of
any astrophysical system. However, chemical evolution models indicate that in
the solar neighborhood the time at which the Fe production from SNe Ia
starts to become important is $\sim 1$Gyr after the beginning of star
formation (Matteucci \& Greggio 1986; MR01), well reproducing the
$[\alpha /Fe]$ values as a function of the $[Fe/H]$ ratio observed in
Galactic field stars. Moreover, MVP06 have shown that a large fraction
of SNe Ia should arise from systems exploding on timescales on the
order of 40Myr, to explain the Type Ia SN rates observed in radio-loud
elliptical galaxies. The data observed by Dahlen et al. (2004) were the
first collected at redshifts $>1$ for the Type Ia SN rate and are
likely to represent lower limits to the actual values (and are still
uncertain). However, recently Kuznetsova et al. (2008) presented a new
measurement of the Type Ia rate up to redshift of 1.7, using two
samples collected by the Hubble Space Telescope. They applied a novel
technique for identifying Type Ia SNe, based on a Bayesian probability
approach, and their results (open circles) appear to
be slightly larger than those of Dahlen et al. (2004).  In this way,
the agreement between the predictions of the models presented here and
the observations improves and there is no more need to assume very
large delay times.  Moreover, it is very important to note that the
assumed star formation history dominates over the assumed DTD and that
one can obtain a cosmic SNIa rate bending down at z$\sim$2 with
a drcreasing SFRd and any DTD.  Therefore, the suggestion of the
long delay time for the SNIa rate is not a robust conclusion anyway.

In Figs. 11, 12, 13 and 14 we show the results obtained by assuming a
different epoch of galaxy formation, $z_{f}=10$. The values of the
realization probability $A_{Ia}$ are the same assumed in the case with
$z_{f}=6$ (as labeled in the figures).  We note in these
figures that the models with $z_f=10$ , computed in the framework of a decreasing SFRd with redshift,
show a SNeIa peak at a slightly larger $z$ compared to the models with
$z_f = 6$, and that all the models show larger SNIa rates at z $\sim$
2.  The behaviour of the models in the redshift interval 0 $< z <$ 1
(the range better covered by observations) does not vary appreciably
when changing the redshift of galaxy formation. Finally, in Fig. 15 we show the results
obtained with the cosmic SFRd of Cole et al. (2001) coupled with all the explored DTDs.
Here the agreement looks good for almost all the DTDs, with perhaps the exception of the DTD relative to the 
close DD model, thus confirming the conclusion that it is very difficult to select a DTD from the observed
cosmic Type Ia SN rate.

\begin{figure}
     \centering
     \includegraphics[width=8.5cm]{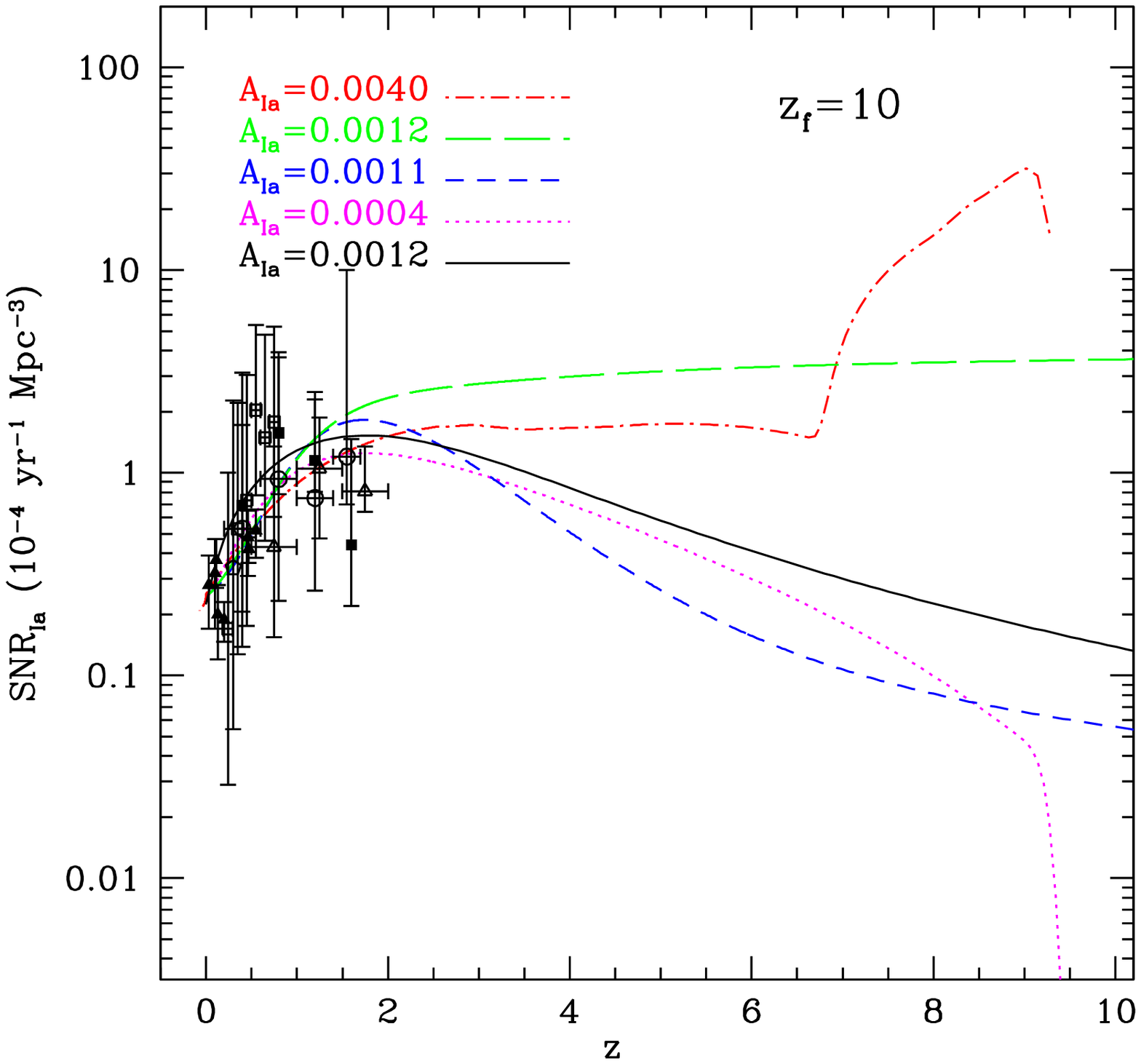}
     \caption{The cosmic SN Ia rate density for the MVP06 DTD assuming $z_f=10$ for the redshift of galaxy formation. The models, units and data are the same as in Fig. 7. \label{SNRcomp1}}
\end{figure}
\begin{figure}
     \centering
     \includegraphics[width=8.5cm]{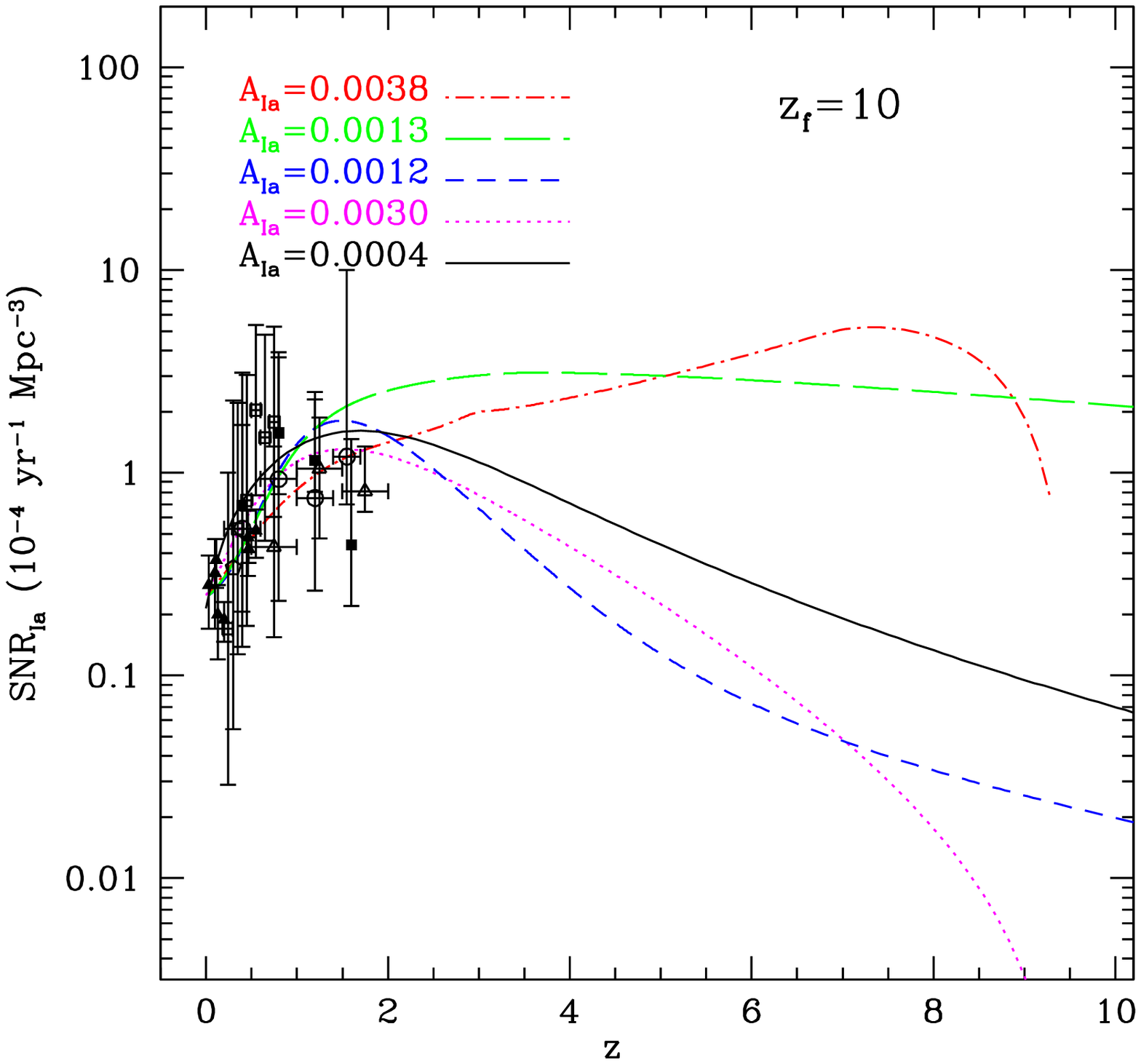}
     \caption{The cosmic SN Ia rate density for the MR01 DTD assuming $z_f=10$  for the redshift of galaxy formation. The models, units and data are the same as in Fig. 7. \label{SNRcomp1}}
\end{figure}
\begin{figure}
     \centering
     \includegraphics[width=8.5cm]{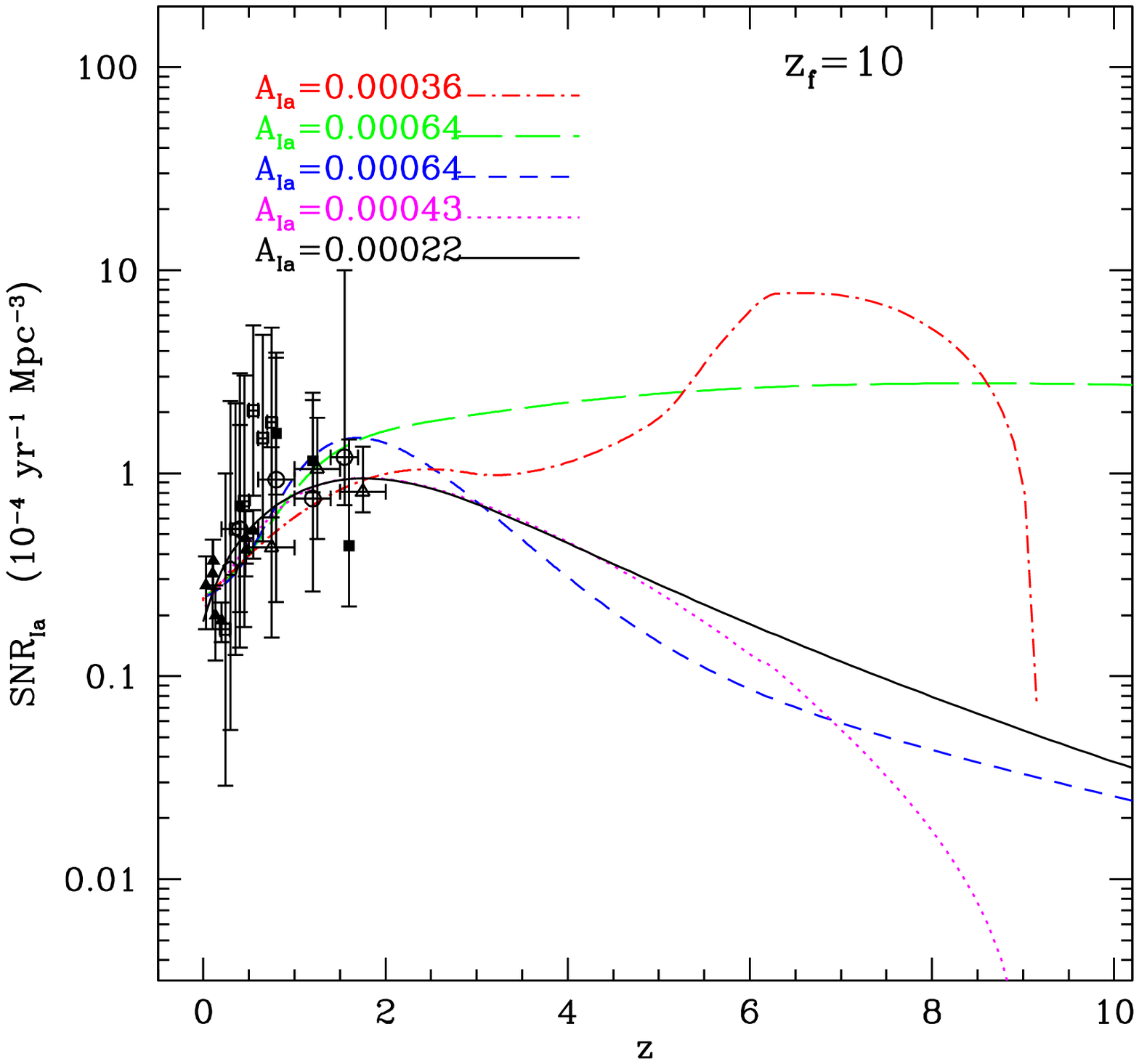}
     \caption{The cosmic SN Ia rate density for the wide DD DTD assuming $z_f=10$  for the redshift of galaxy formation. The models, units and data are the same as in Fig. 7.  \label{SNRcomp1}}
\end{figure}
\begin{figure}
     \centering
     \includegraphics[width=8.5cm]{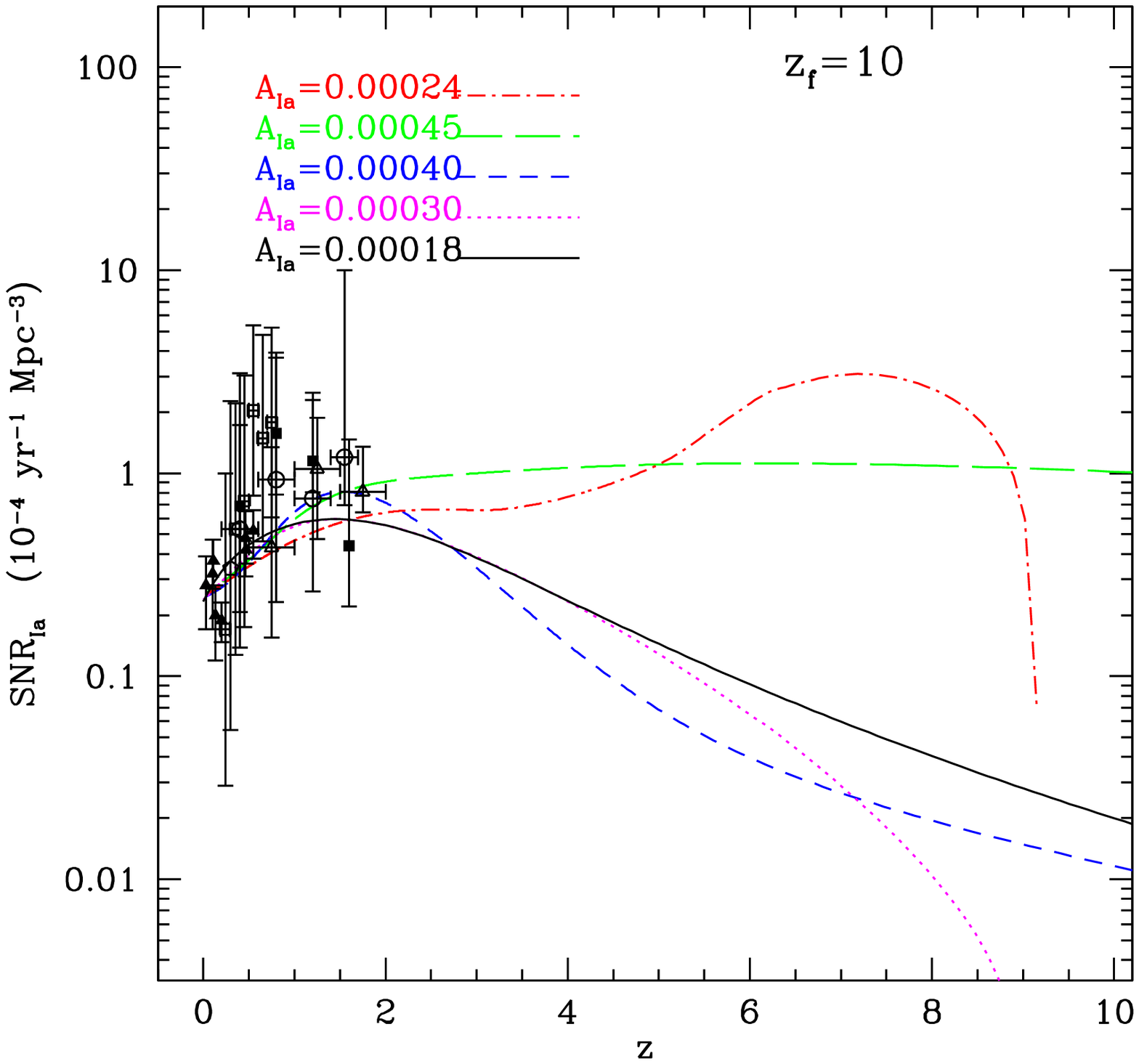}
     \caption{The cosmic SN Ia rate density for the close DD DTD assuming $z_f=10$ for the redshift of galaxy formation. The models, units and data are the same as in Fig. 7. \label{SNRcomp1}}
\end{figure}
\begin{figure}
     \centering
     \includegraphics[width=8.5cm]{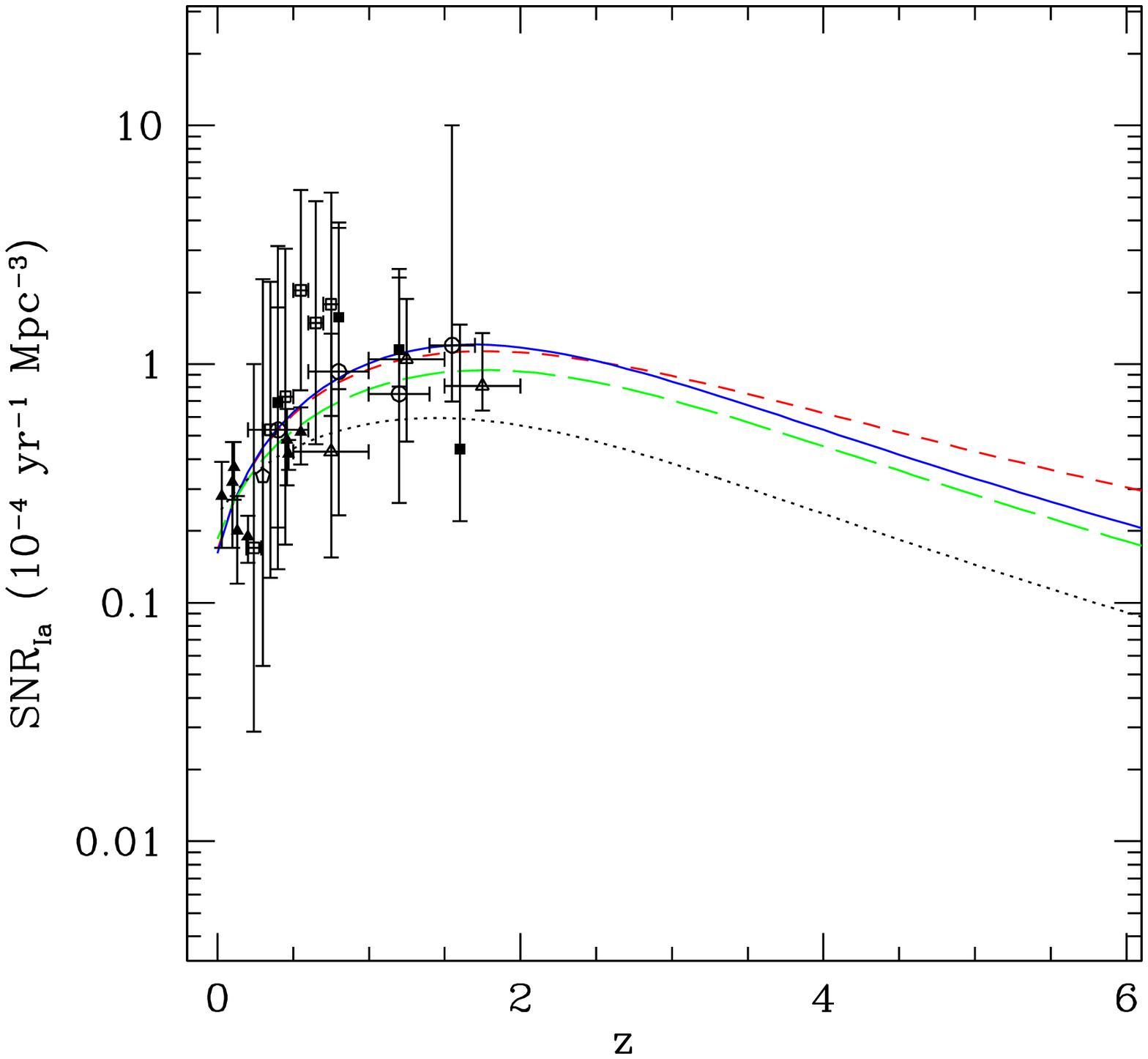} 
     \caption{The predicted cosmic SN Ia rates adopting the SFR density from Cole et al. (2001). The short-dashed line is for the MVP06 DTD, the solid line is for the MR01 DTD while the dashed and dotted lines are for the wide DD and close DD DTDs, respectively (G05). The figure underlines the differences due to the choice of the DTDs. The data are the same as in the previous Figures. Note that this diagram is independent of the redshift of galaxy formation, since the SFR density adopted does not depend on $z_f$. \, \, \, \, \, \, \, \, \, \, \, \, \, \, \, \, \, \, \, \, \, \, \, \, \, \, \, \, \label{SNRcomp1}}    
\end{figure}
\section{Conclusions}
We calculated the cosmic SN Ia rate densities (i.e. the rate per unit comoving volume) and the rate of the explosion of SNe Ia in typical elliptical galaxies adopting the formalism proposed by G05. This formulation rests upon the definition of the SN Ia rate following an instantaneous burst of star formation as a function of the time elapsed from the birth of the progenitor system to its explosion as a Type Ia SN (i.e. the delay time). This function has been termed as DTD and accounts for the SNe Ia progenitor scenarios and for the initial mass function (IMF). Different DTDs and histories of SF have been considered. In all cases, a Salpeter IMF  was adopted. For all galaxies it was assumed a galaxy  formation epoch at redshift $z_{f}=6$. The reason for chosing $z_f=6$ is given by the fact that the SFR density has been measured up to this redshift (Cole et al. 2001).
We have also shown the results obtained assuming $z_{f}=10$ as epoch of galaxy formation.
The evolution of the cosmic SN rate with redshift contains, in principle, unique information on the star formation history of the Universe, the IMF of stars, and the Type Ia SN progenitors. These are essential ingredients for understanding galaxy formation, cosmic chemical evolution, and the mechanisms which determined the efficiency of the conversion of gas into stars in galaxies at various epochs (e.g. Madau et al. 1996; Madau, Pozzetti, \& Dickinson 1998; Renzini 1997). 
We computed the cosmic Type Ia SN rates for different cosmic SFR histories. Our main result is the prediction of the expected number of explosions of Type Ia SNe at high redshift ($z\geq 2$). 

Here we summarize our  conclusions in detail:
\begin{itemize}
\item we analyzed the effects of the various parameters entering the computation of the SN Ia rate and concluded that the realization probability $A_{Ia}$ (the actual fraction of systems which are able to give rise to a Type Ia SN) should be in the range $10^{-4}-10^{-3}$, where the lower value is appropriate for the DD model. This quantity, in all models, is a free parameter and it was chosen by reproducing the present time Type Ia SN rate in galaxies. 
Other parameters also play a key role in the computation of the Type Ia SN rate and these are: \emph{i)} the mass range for the secondaries; \emph{ii)} the minimum mass for the primaries; \emph{iii)} the efficiency of the accretion and \emph{iv)} the distribution of the separations at birth.
Actually, all the different channels could contribute to the SN Ia events, each with its own probability ($A_{Ia}$). 
\item We computed the Type Ia SN rate for specific elliptical galaxies of different initial luminous masses ($10^{10}$M$_{\odot}$, $10^{11}$M$_{\odot}$, $10^{12}$M$_{\odot}$) for all the studied DTDs and for the SFRs suggested by chemical evolution models which best reproduce the characteristics of local ellipticals (Pipino \& Matteucci 2004). All the DTDs predict an early maximum in the Type Ia SN rate, between 0.3 and 1 Gyr, according to the galaxy mass. In particular, the maximum predicted at about 0.3 Gyr corresponds to a typical elliptical of initial mass of $10^{12}$M$_{\odot}$ while, for an elliptical of initial mass of $10^{10}$M$_{\odot}$ the peak is at about 1 Gyr. 
The dependence of the maximum on the initial galactic mass is due to the efficiency of star formation, which is assumed to be higher for the most massive ellipticals . 
We have also considered the predicted Type Ia SN rates per unit mass (SNuM) at the present time vs. $(B-K)$ color relations for the three different ellipticals. We found that the bluer is the color (and hence the lower is the galactic mass), the higher are the predicted SN Ia rates in SNuM at the present time. 
Unfortunately, a real comparison with data is not possible since the data for ellipticals of different masses are not available.
\item We computed the cosmic Type Ia SN rate in a unitary volume of Universe by adopting different cosmic SFRds, as predicted by both models and observations (Hopkins \& Beacom, 2006). We found that the SF history largely dominates over the assumed DTD in the calculation of the Type Ia SN rate. Therefore, unless we know the cosmic star formation history, we cannot safely decide which DTD is better on the basis of the observed cosmic SN Ia rate, in agreement with previous works (see also Forster et al. 2006; Blanc \& Greggio 2008, Botticella et al. 2008).
In particular, it is not possible to predict the delay time for the explosion of SNe Ia on the basis of the cosmic star formation rate, since for example, in the hierarchical framework, it is predicted a decrease of the cosmic Type Ia SN rate for $z>1$ irrespective of the chosen DTD, even for that of Mannucci et al. (2005; 2006) DTD,  where the fraction of prompt Type Ia SNe is 50$\%$!

\item The cosmic Type Ia SN rates for different SFRds (increasing, decreasing, constant with redshift) differ mostly for redshift $z>1$. We compared our results with the available data, although the highest redshift points are still very uncertain. Therefore, it was not possible to decide which scenario should be preferred, but high redshift data will greatly help to draw conclusions on this point. In particular, we predict that the Type Ia SN rate for $z>2$ should be very high if the monolithic scheme for the formation of ellipticals is assumed and even higher if combined with the empirical DTD of Mannucci et al. (2005), as opposed to models with decreasing or constant SFRds, which all predict a significant drop of the Type Ia SN rate for any chosen DTD. Future observations with JWST of high redshift SNe will help in sheding light on this subject.

\end{itemize}
\section*{Acknowledgments}

We wish to thank Antonio Pipino for the help and collaboration.\\ 
We also thank  I.J. Danziger, Avishay Gal-Yam, Dan Maoz and  Filippo Mannucci for 
valuable suggestions and comments.
We also thank an anonymous referee, whose comments and suggestions greatly helped to improve this paper



\label{lastpage}

\end{document}